\documentclass[twocolumn,amsmath,pra]{revtex4-1}
\usepackage{graphicx}
\usepackage{color}
\usepackage{todonotes}

\renewcommand{\vec}[1]{\mathbf{#1}}

\begin{document}

\title{Stability and Dynamics of Cross Solitons in Harmonically Confined Bose-Einstein Condensates}

\author{Tadhg Morgan}
\email{tmorgan@phys.ucc.ie}
\homepage{http://groups.oist.jp/qsu}
\affiliation{Quantum Systems Unit, OIST Graduate University, Okinawa, Japan}
\affiliation{University College Cork, Cork, Ireland}

\author{Thomas Busch}
\affiliation{Quantum Systems Unit, OIST Graduate University, Okinawa, Japan}

\begin{abstract}
We examine the stability and dynamics of a family of crossed dark solitons in a harmonically confined Bose-Einstein condensate in two dimensions. Working in a regime where the fundamental {\it snake instability} is suppressed, we show the existence of an
instability which leads to an interesting collapse and revival of the initial state for the fundamental case of two crossed solitons. The instability originates from the singular point where the solitons cross, and we characterise it by examining the Bogoliubov spectrum. Finally, we extend the treatment to systems of  higher symmetry.
\end{abstract}

\maketitle

\section{Introduction}
\label{sec:Introduction}
Bose-Einstein condensates (BECs) of weakly interacting, ultracold atomic gases provide highly controllable systems in which one can explore non-linear properties of matter waves \cite{Stringari:03}. One effect stemming from the balance between nonlinear and dispersive effects is the existence of solitary matter waves, or solitons, and these have in recent years been subject to extensive theoretical \cite{Reinhardt:97,Busch:00,Parker:09,Gardiner:07} and experimental investigation \cite{Strecker:02,Burger:99,Denschlag:00,Dutton:01,Anderson:01,Khaykovich:02,Becker:08,Zwierlein:13}. Solitons can be both, bright and dark, depending on the type of nonlinearity in the system and their primary attribute is propagation without dispersion. 

Single solitons and soliton-soliton collisions in weakly interacting condensates in harmonic traps have been thoroughly examined in recent years \cite{Stellmer:08,Gardiner:07}. However, due to the known instabilities in higher dimensions \cite{Brand:02,Toikka:13}, most work has concentrated on one-dimensional (1D) and quasi-1D (ring) geometries \cite{Reinhardt:97,Busch:00,Parker:09, Gardiner:07,Liu:05,Walczak:11}. At the same time there have been a number of studies of higher-dimensional solitons in dipolar condensates \cite{Pedri:05} and periodic potentials \cite{Baizakov:03,Baizakov:04}, where different stability properties can be found. Here we extend the discussion of solitonic solutions in weakly interacting BECs to multidimensional setups and discuss the appearance of a new type of instability.

\begin{figure}[t]
\includegraphics[width=0.95\columnwidth]{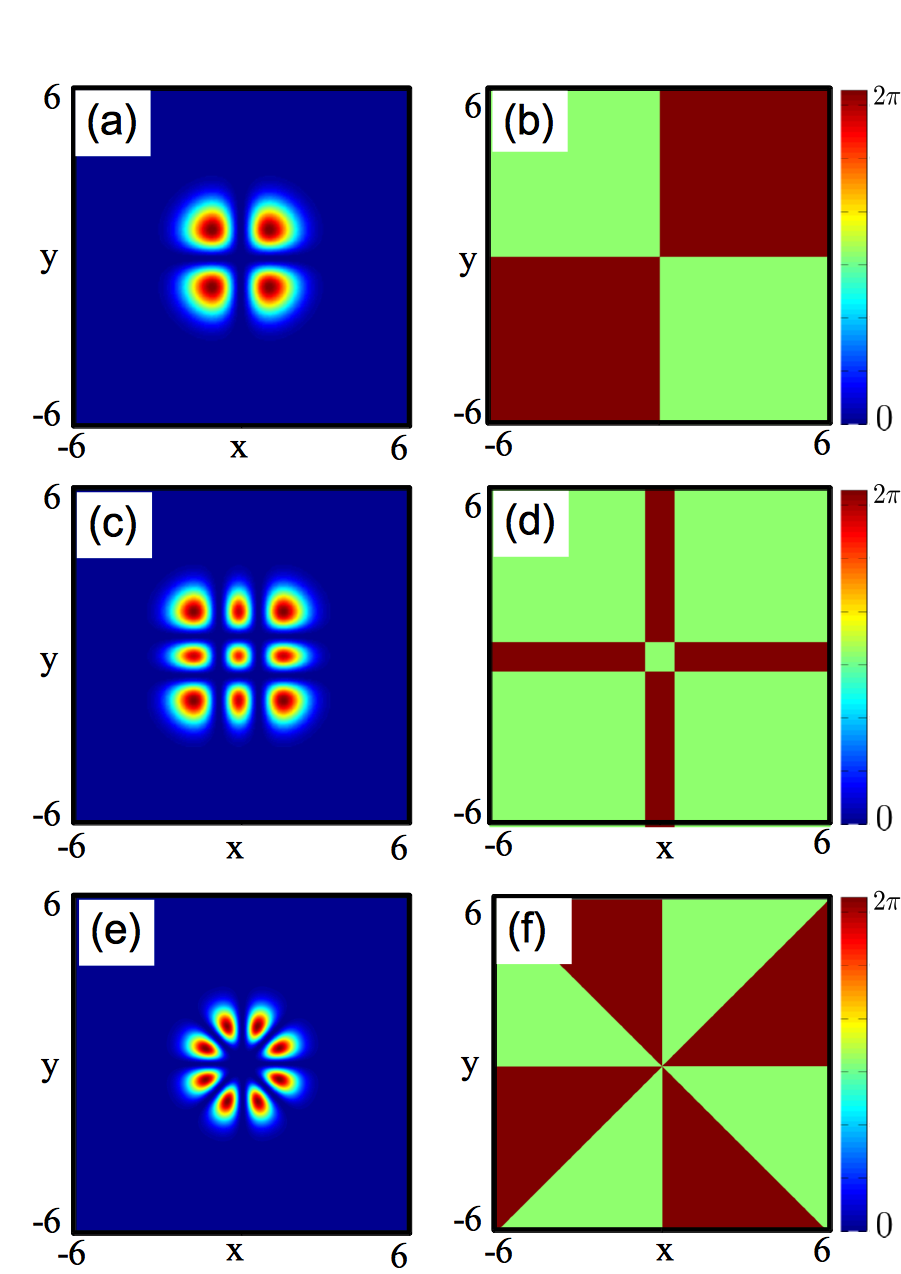}
\caption{\label{fig:Family} (Color online) Density (left column) and phase (right column) of (a) and (b) the cross soliton, (c) and (d) the double-cross soliton, and (e) and (f) a star soliton. For all cases the interaction strength $\tilde{g}=11$ [see Eq.~\ref{eq:GP}].}
\end{figure}

The two-dimensional soliton states which we examine in this work are part of a family of so-called dark solitons, which are characterised by a phase profile where each area of distinct phase differs from all other neighbouring areas by a difference of $\pi$. Such a profile produces a density dip across each phase jump, which is stable as long as the phase difference is maintained.  The configurations which we consider here are comprised of several dark soliton lines in two dimensions, which overlay and intersect with each other at different angles. In particular, we focus on the arrangements shown in Fig.~\ref{fig:Family}, which represent the most fundamental and symmetric geometries \cite{Xu:09,Feder:00}.

The basic instability of a single dark soliton in two dimensions is to eventually decay into a number of quantized vortices through what is called the {\it snake instability} \cite{Brand:02,Toikka:13}. This instability is due to the extension of the soliton into the direction orthogonal to the phase jump and causes it to bend (or {\it snake}) until the curvature is large enough to lead to a break-up into vortex-antivortex pairs. In inhomogeneous systems, however, the snake instability can be suppressed by reducing the width of the trap until the lowest mode of the snake instability is no longer allowed \cite{Brand:02}.

This is the regime we work in here, as it will allow us to clearly identify and describe possible new instabilities originating from the singular points where the solitons cross  [see Figs.~\ref{fig:Family}(b), \ref{fig:Family}(d), and \ref{fig:Family}(f)]. For this we numerically integrate the Gross-Pitaevskii equation of the system and find an instability corresponding to distinct areas of equal phase connecting across the singular points. Due to the finite size of our system, we also observe a disconnection and almost perfect revival of the initial state, after which the process happens again. To further understand the nature of the instability we make use of the well known Bogoliubov-de Gennes equations to obtain the linear excitation spectrum of the system.

The layout of this paper is as follows. In Sec.~\ref{sec:Initial States} we outline the nature and creation of the initial states of the solitons shown in Fig.~\ref{fig:Family}. In Sec.~\ref{sec:Dynamical Instability and Bogoliubov Analysis} we numerically time evolve the initial state of the cross soliton to study the dynamics arising from its instability and calculate the Bogoliubov spectrum and the associated eigenfunctions. We show that the Bogoliubov spectrum contains a complex eigenfrequency, which can be connected to the observed dynamical instability. In Sec.~\ref{sec:Higher Order Symmetry} we discuss the stability and dynamics of higher order solitons such as the double-cross soliton [see Figs.~\ref{fig:Family}(c) and \ref{fig:Family}(d)] and the star soliton [see Figs.~\ref{fig:Family}(e) and \ref{fig:Family}(f)]. Finally, in Sec.~\ref{sec:Conclusion} we conclude.

\section{Initial States}
\label{sec:Initial States}
In this section we briefly introduce and discuss the initial states of the solitonic systems we wish to study. As we consider inhomogeneous Bose-Einstein condensates of neutral, atomic gases, the condensate wave function $\psi$ will be described at any point in time by the time-dependent Gross-Pitaevskii equation \cite{Stringari:03}. For numerical tractability we restrict ourselves to a two-dimensional (2D) BEC of $N$ atoms of mass $m$ confined to an isotropic harmonic oscillator with trapping frequency $\omega_T$ and rescale our coordinates to make them dimensionless via \cite{Bao:03}
\begin{align}
   \tilde{t}=\omega_T t, \quad \tilde{x}=\frac{x}{a_0}, \quad \tilde{y}=\frac{y}{a_0},\quad
\label{eq:Scaling}
\end{align}
where $a_0=\sqrt{\hbar/\omega_T m}$ is the length of the harmonic oscillator ground state and all energies are in units of $\hbar \omega_T$. For ease of notation, we will in the following drop the tilde again. The dimensionless Gross-Pitaevskii equation can then be written as
\begin{equation}
\label{eq:GP}
  i \frac{\partial \psi}{\partial t}=\left[-\frac{1}{2} \nabla^2 + \frac{1}{2}(x^2+y^2) + \tilde{g} |\psi|^2 \right]\psi,
\end{equation}
where the nonlinear interaction strength is given by $\tilde{g}=\frac{4 \pi a_s N}{a_0} \sqrt{\frac{\gamma_z}{2 \pi}}$. Here $a_s$ is the $s$-wave scattering length of the atomic species and $\gamma_z=\omega_z / \omega_T$, with $\omega_z$ being the trapping frequency in the $z$ direction. To numerically generate the initial states shown in Fig.~\ref{fig:Family}, we evolve an initial wave function in imaginary time using the fast Fourier transform (FFT) split-operator method \cite{Fleck:76} under the condition that the desired phase pattern is maintained. The FFT-split-operator method is also used for real-time evolution of the Gross-Pitaevskii equation (\ref{eq:GP}). 

The fundamental example of a two-dimensional soliton structure is given by the cross soliton, shown in Figs.~\ref{fig:Family}(a) and \ref{fig:Family}(b). It consists four symmetric lobes separated by a density dip, and the phases between neighbouring lobes are arranged to differ by the required factor of $\pi$. While the angle between the solitons can, in principle, take any value, we concentrate here on the perpendicular setting as it allows us to clearly identify the nature of the instability. Note that the dark-soliton lines in this state are in the radial direction and therefore are not subject to the well-known oscillation instability \cite{Busch:00}.

A higher-order state of the family of crossed solitons is the so-called double-cross soliton, which is shown in Figs. \ref{fig:Family}(c) and \ref{fig:Family}(d). It consists of two pairs of perpendicularly crossed density dips, leading to nine separated density areas with appropriate phases between them. Due to the presence of the external, harmonic potential, not all areas are equally populated, and the dark-soliton lines are no longer radial lines. This leads to small oscillations of the solitons in the potential \cite{Busch:00}, but we find them to not influence the newly forming instabilities.

The final system we will investigate is the so-called star soliton, shown in Figs.~\ref{fig:Family}(e) and \ref{fig:Family}(f). It is an extension of the cross soliton described above, but instead of four lobes the condensate is split into eight with appropriate phases between them. All soliton lines connect in the centre, which means that they are radial lines and no extra oscillations are to be expected. Again, while, in principle, all angles between the intersecting solitons can be considered, we focus here on the symmetric setting where all angles are chosen to be $\pi/4$.

\section{Dynamical Instability and Bogoliubov Analysis}
\label{sec:Dynamical Instability and Bogoliubov Analysis}

In order to determine the stability of the states described above, we will carry out a fully two-dimensional integration of the Gross-Pitaevskii equation and examine the eigenspectrum obtained from a Bogoliubov analysis. For the latter one has to solve the Bogoliubov-de Gennes equations \cite{Bogoliubov:47,Stringari:03}
\begin{align}
  L \vec{u_j} - \tilde{g} \psi^2 \vec{v_j} &= \omega_j \vec{u_j},\\
  L \vec{v_j} - \tilde{g} \psi^{*2} \vec{u_j} &= -\omega_j \vec{v_j},  
\end{align}
where $L=-\frac{1}{2}\nabla^2 + \frac{1}{2} (x^2+y^2) + 2 \tilde{g} \vert\psi\vert^2 - \mu$ and $\mu$ is the chemical potential. Examining the spectrum of eigenfrequencies $\omega_j$ and corresponding eigenvectors $\vec{u_j}$ and $\vec{v_j}$  provides information about the stability properties of the state $\psi$. A small positive $\omega_j$ with a positive norm $n_j=\int(\vert \vec{u_j} \vert ^2 -\vert \vec{v_j} \vert ^2) d\vec{r}$ corresponds to small oscillations about the state and indicates stability. A mode with negative $\omega_j$ with a positive norm $n_j$ is called an anomalous mode and indicates that the initial state will continuously transform into a lower-energy state. Finally, complex and purely imaginary eigenfrequencies $\omega_j$ with $n_j=0$ indicate the presence of a dynamical instability \cite{Stringari:03}.

To be able to identify potential instabilities cleanly, we must work in a regime where the snake instability for line-solitons is suppressed. As shown by Brand and Reinhardt \cite{Brand:02}, this can be achieved by reducing the transverse width of the soliton, and for a condensate consisting of a large number of repulsively interacting atoms the condensate width can be determined in the Thomas-Fermi approximation as \cite{Bao:03}
\begin{equation}
R=\frac{1}{\sqrt{2}}\left( \frac{15 \tilde{g}\gamma_z}{4 \pi} \right ) ^{1/5}.
\end{equation} 
One can see that the size of the condensate, and hence the transversal width of an embedded soliton, can be reduced by decreasing the nonlinear interaction constant $\tilde{g}$.

Calculating the Bogoliubov spectrum for a single dark soliton, we find that it becomes completely real and positive once $\tilde{g}\le 19$, which corresponds to the point beyond which the snake instability is suppressed \cite{remark1}. We have confirmed this by direct numerical time evolution and also checked that around this value the decay into vortices is absent in the multisoliton systems shown in Fig.~\ref{fig:Family}. To ensure that we are far from the snake instability regime for all cases we therefore choose $\tilde{g}=11$ in all numerical simulations.

\begin{figure}[t]
\includegraphics[width=0.95\columnwidth]{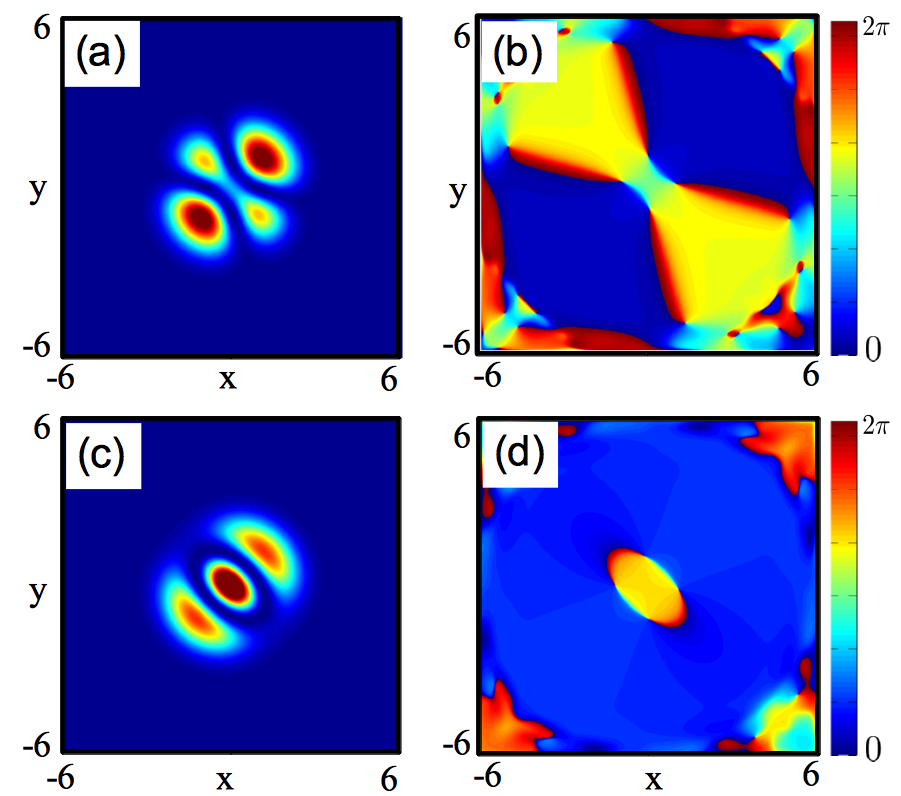}
\caption{\label{fig:Initial}(Color online) The densities (left column) and phases (right column) of the cross soliton as it begins to collapse at $t=128$ (top row) and after full collapse at $t=138$ (bottom row). }
\end{figure}

Although the snake instability is suppressed, real-time evolution of the cross soliton in this regime reveals the existence of another dynamic instability. After a period where the initial state is stationary, a series of repeating collapse and revival events sets in, which are driven by one of the pairs of lobes of equal phase connecting and collapsing to the centre of the trap with the other pair of lobes surrounding it [see Figs.~\ref{fig:Initial}(a) and \ref{fig:Initial}(b)] \cite{remark2}. The density and phase of the fully collapsed cross soliton are shown in Figs.~\ref{fig:Initial}(c) and \ref{fig:Initial}(d), and the state has the form of two curved line solitons. In fact, by examining the low-density areas it can be seen that the two outer parts connect to form a highly nonsymmetric ring soliton. This collapsed state is relatively short-lived and evolves, due to the finite size of the system, back into the initial state seen before the collapse \cite{Parker:03}. If we further evolve the state in time we see another collapse and revival, but this time the other pair of lobes connects and collapses to the centre.

\begin{figure}[t]
\includegraphics[width=0.95\columnwidth]{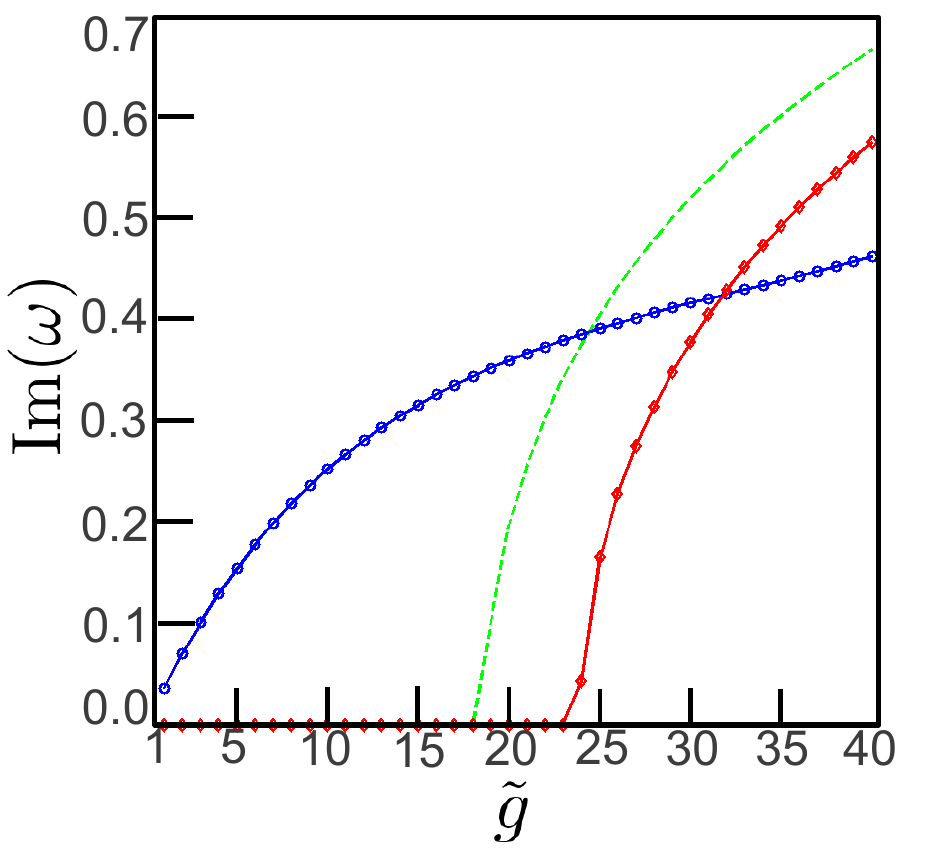}
\caption{\label{fig:Spec}(Color online) The imaginary modes of the Bogoliubov spectrum of the cross soliton. The red line with diamonds corresponds to the snake instability, and the blue line with circles corresponds to the new cross soliton instability. The green dashed line is included for reference and corresponds to the snake mode of a single dark-line soliton.}
\end{figure}

In Fig.~\ref{fig:Spec} we show the  imaginary frequencies of the Bogoliubov spectrum for the cross soliton as a function of the interaction strength $\tilde{g}$. These correspond to the instability identified above (blue line with circles) and the snake instability (red line with diamonds).  
One can see that the snake instability only sets in once the repulsive interaction has increased the size of the condensate beyond the critical width and a finite region exists in which the instability discussed above sets in before the snake instability. Note that the line corresponding to the snake instability is actually doubly degenerate, corresponding to an instability in each of the crossed solitons.

For reference we have also included the sole imaginary Bogoliubov mode of the single dark soliton, which corresponds to the snake instability (green dashed line). The shift between this line and the one for the crossed soliton indicates that the crossing point leads to a certain increase in stability against {\it snaking}.

\begin{figure}[t]
\includegraphics[width=0.95\columnwidth]{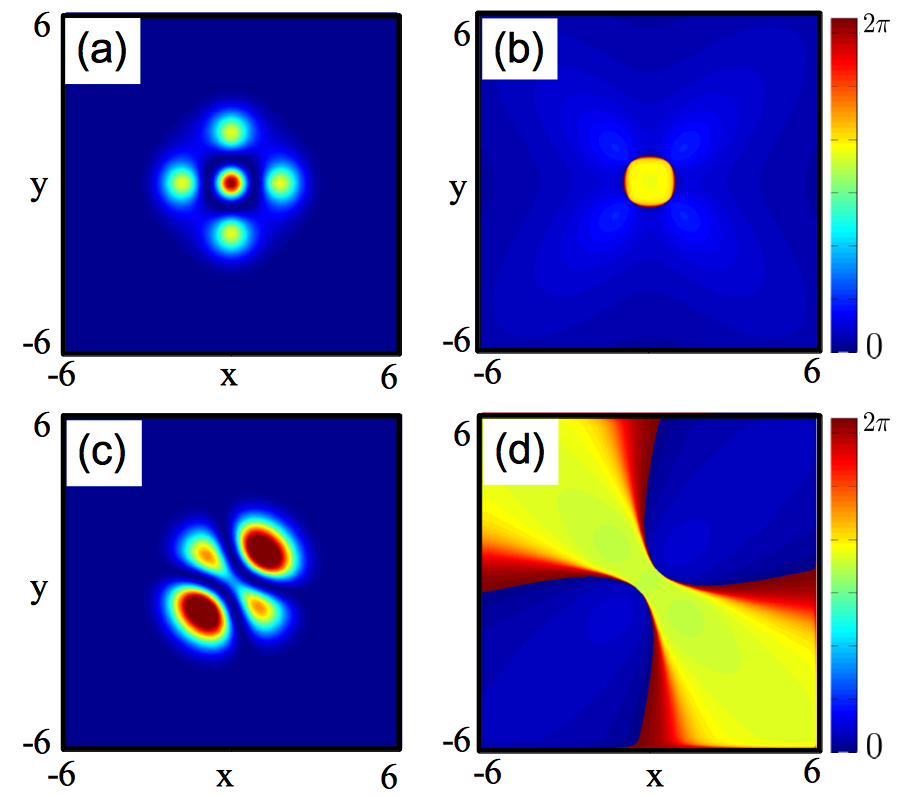}
\caption{\label{fig:Modes} (Color online) (a) The density and (b) phase of the unstable Bogoliubov mode with eigenfrequency $\omega_1=0.2673i$ for the cross soliton. (c) The density and (d) phase of $\Psi$ for $\alpha=25$. Note that the color scales of (a) and (b) are not equal.}
\end{figure}

At the value of $\tilde{g}=11$ considered in our numerical simulations, only one imaginary frequency mode with $\omega_1=0.2673i$ exists, the density and phase of which are shown in Figs.~\ref{fig:Modes}(a) and \ref{fig:Modes}(b), respectively. To show that this mode is indeed the mode responsible for the instability we observe, we examine the density and phase of a state that is created by linearly combining the initial cross soliton state $\psi$ and the unstable mode,
\begin{equation}
  \Psi= \psi+ \alpha \vec{u_j}.
  \label{eq:Combo}
\end{equation}
As the form of the Bogoliubov modes is given by $\vec{u_j} e^{-i \omega_j t}$, over a short period of time, an imaginary frequency will cause the amplitude of the mode to increase exponentially before, due to interference from waves reflected on the boundary of our system, it reduces again. By using a nonzero $\alpha$ parameter in (\ref{eq:Combo}), we can therefore approximate the influence of the mode on the initial state, and as can be seen in Figs.~\ref{fig:Modes}(c) and \ref{fig:Modes}(d), for $\alpha=25$, the density and phase of $\Psi$ are very close to those obtained numerically at $t=132$ [Figs.~\ref{fig:Initial}(a) and \ref{fig:Initial}(b)].

\begin{figure}[t]
\includegraphics[width=0.95\columnwidth]{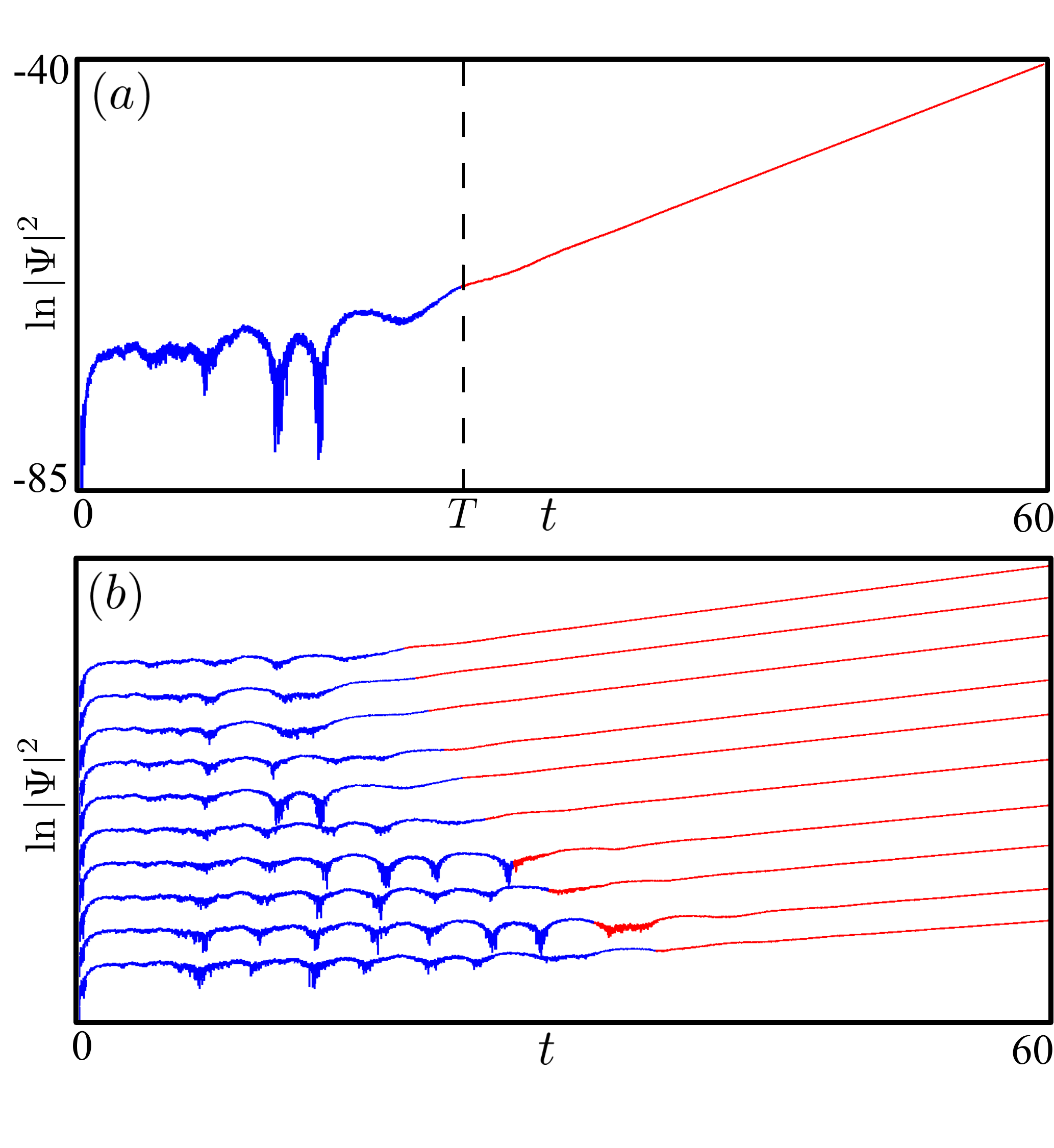}
\caption{\label{fig:Instability_Time} (Color online) (a) Density at the trap centre as a function of time for a condensate with $\tilde{g}=11$. The time for the onset of the instability predicted by the Bogoliubov analysis is indicated by the dashed line and by the change of colour of the graph from blue (dark gray) to red (light gray). (b) Same as in (a) for different non-linearities ranging from $\tilde{g}=6$ to $\tilde{g}=15$ in steps of $1$ from below. The individual curves are offset for clarity.}
\end{figure}

The time for the onset of the instability can be predicted from the eigenfrequency of the unstable mode as $T\approx\frac{2 \pi}{\text{Im}(\omega_1)}$. This can be seen in Fig.~\ref{fig:Instability_Time}(a), where we show the density at the centre of the trap (the point of the singularity) as a function of time. Initially, the system performs small oscillations which, around the predicted time of the instability $T$, turn into an exponential increase in the density. This ultimately leads to the complete collapse of the cross soliton. 

In Fig.~\ref{fig:Instability_Time}(b) we repeat this analysis over a wide range of values for $\tilde{g}$. The time for the instability $T$ is represented by the change of the line colour from blue (dark gray) to red (light gray), and one can see that the effect is consistent over the whole range. Furthermore, these data also show that increasing the strength of interaction leads to an earlier onset of instability in the system.
\section{Higher-Order Symmetry}
\label{sec:Higher Order Symmetry}
In this section we will extend the discussion of stability presented for the cross soliton to the higher-order structures shown in Fig.~\ref{fig:Family}. However, before we can do so, we briefly need to address the stability of our numerical approach. One cause for concern about the validity of the presented Bogoliubov analysis is the influence of the square numerical grid used in the generation of the initial state. Any dark-soliton line not on axis with the grid can only be approximated and will suffer from spatial aliasing, which in turn leads to a numerical effect on the instability time scales. We have investigated this issue thoroughly and found that this problem is absent for a dark-soliton line at an angle of  $\pi/4$ to the numerical grid by comparing the results obtained for a cross soliton at an angle of $\pi/4$ with the ones obtained for the structure lying along the grid axes. 

Therefore, the first higher-order soliton we study is the double-cross soliton, whose composite dark-soliton lines are on axis with the numerical grid. The second soliton structure is the star soliton, which contains two dark-soliton lines that are on axis with the numerical grid and two at an angle of $\pi/4$ to the numerical grid.
\subsection{Double-Cross Soliton}
\label{sec:DCS}
For the double-cross soliton, the density and phase are shown in Figs.~\ref{fig:Family}(c) and \ref{fig:Family}(d). Time evolution of this state reveals that it is, much like the cross soliton, quasistable for a certain period before the instability sets in. In this period of quasistability, however, the soliton lines are subject to the oscillation instability in inhomogeneous potentials \cite{Busch:00}, but the amplitudes gained are small and do not affect the onset of the collapse instability.

\begin{figure}[tb]
\includegraphics[width=\columnwidth]{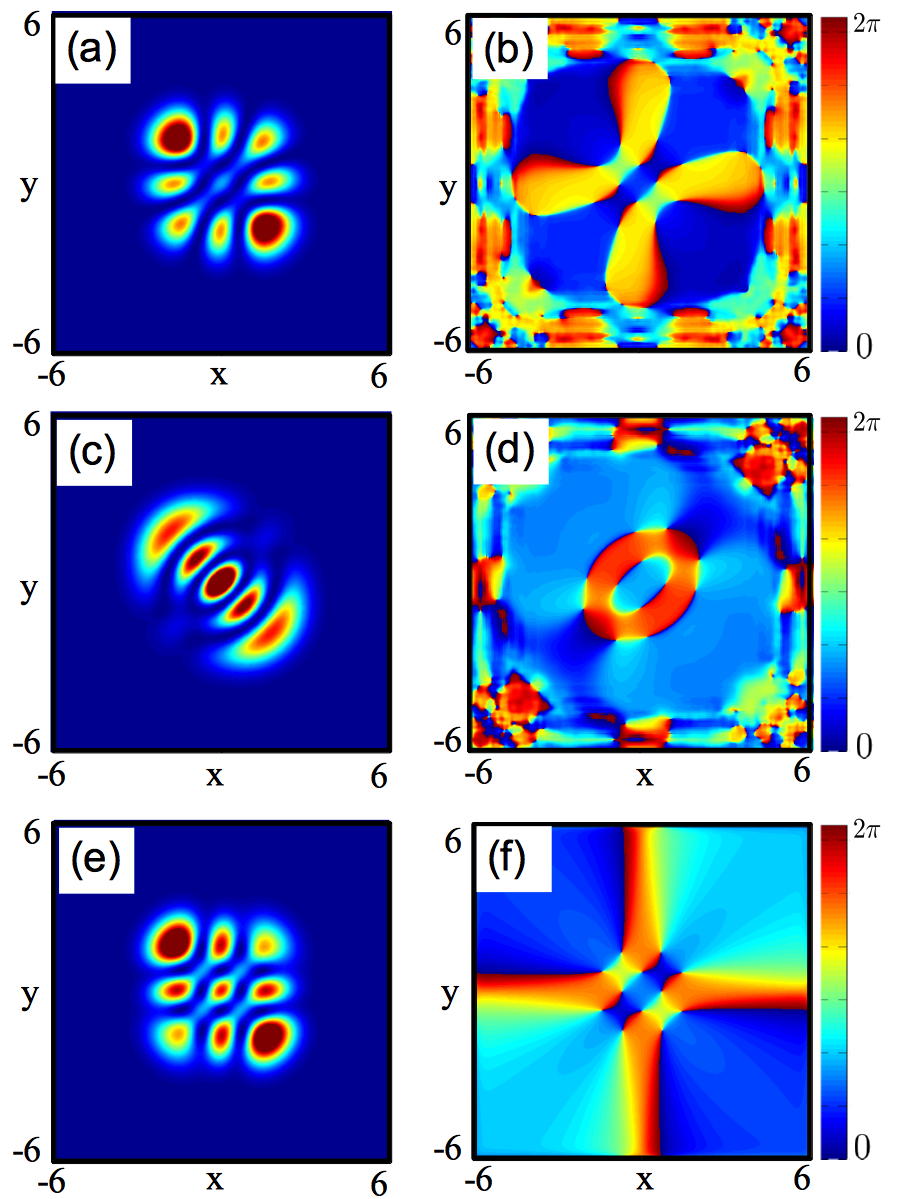}
\caption{\label{fig:Double}(Color online) (a) The density and (b) phase of the double-cross soliton in the early stages of the collapse at $t=119$ and (c) and (d) the same for the fully collapsed state at $t=132$, respectively. (e) The density and (f) phase of $\Psi$ for $\alpha=30$.}
\end{figure}

The nature of the collapse of the double-cross soliton is similar to that of the cross soliton, but of a higher order. Like the cross soliton, lobes of equal phase connect and fall into the centre, as shown in Figs.~\ref{fig:Double}(a) and \ref{fig:Double}(b). The density and phase distribution of the fully collapsed double-cross soliton are shown in Figs.~\ref{fig:Double}(c) and \ref{fig:Double}(d), and one can see that the result of the collapse leads to four curved soliton lines. When taking the low-density regions into account and by looking at the phase distribution, one can see that these lines connect to form two concentric, nonsymmetric ring solitons. Unlike the cross soliton, however, we do not observe a revival of the initial state on the time scales we are able to simulate.

The Bogoliubov analysis of the double-cross soliton reveals  one unstable mode with a purely imaginary eigenfrequency $\omega_1=0.2591i$, which is responsible for the observed instability. As for the single soliton, we use the linear combination of the initial state and the unstable mode [see Eq.~\eqref{eq:Combo}] to confirm that this mode corresponds to the observed instability, and Figs.~\ref{fig:Double}(e) and \ref{fig:Double}(f) show a good agreement with the state of the double-cross soliton at $t=119$ for $\alpha=30$.

\begin{figure}[tb]
\includegraphics[width=\columnwidth]{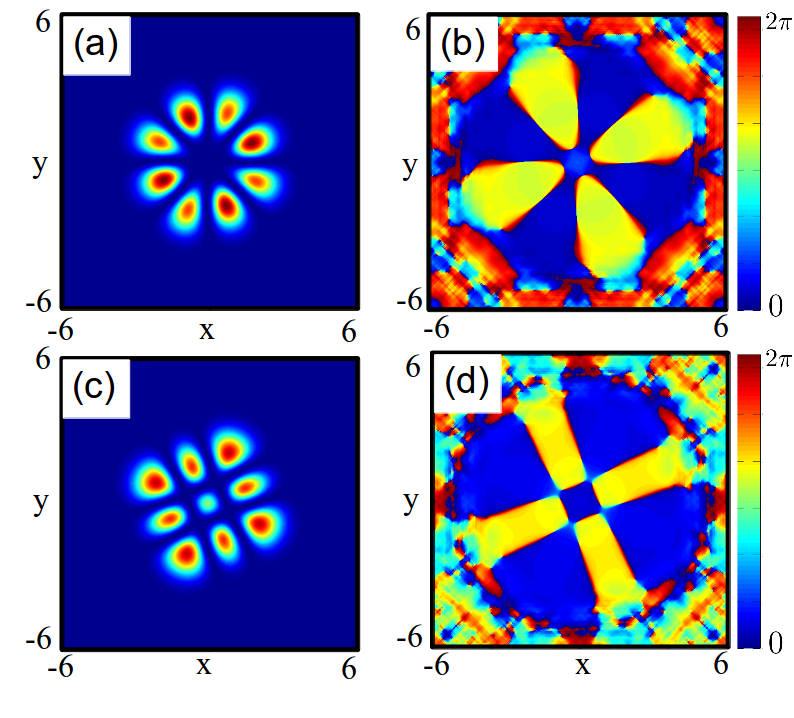}
\caption{\label{fig:Star}(Color online) The density (left column) and phase (right column) of the star soliton during time evolution. (a) and (b) correspond to $t=15$, where the first unstable mode of frequency $\omega_1=0.1211i$ causes slight oscillations in the density distribution between the lobes of the star soliton. (c) and (d) correspond to $t=35$, where the influence of the second unstable mode of frequency $\omega_2=0.0886i$ can be seen.}
\end{figure}

\subsection{Star Soliton}
Unlike the two previous cases, the Bogoliubov spectrum for the star soliton [see Figs.~\ref{fig:Family}(e) and \ref{fig:Family}(f)] reveals two unstable modes with frequencies $\omega_1=0.1211i$ and $\omega_2=0.0886i$. However, due to the increased complexity of the star soliton, producing accurate modes using the Bogoliubov-de Gennes equations is difficult. Therefore, to obtain information about the properties of these instabilities we study the time evolution of the star soliton by numerically integrating the Gross-Pitaevskii equation. 

As indicated by its larger frequency, the influence of the first mode is seen before the second mode during time evolution and corresponds to oscillations originating from lobes of equal phase connecting in the centre [see Figs.~\ref{fig:Star}(a) and \ref{fig:Star}(b)]. This is mode is similar to the one found for the cross soliton and generalises it to the higher order of the star soliton. In contrast, however, we find that the amplitude here is small and the overall star pattern is maintained.
Therefore, it is possible for the second unstable mode to set in at a later time and after the first instability has undergone a couple of oscillatory cycles.

The second mode induces a transformation into a very structured state that is reminiscent of the double-cross soliton, as can be seen in Figs.~\ref{fig:Star}(c) and \ref{fig:Star}(d).  This state, in turn, then decays in the same manner as discussed in Sec.~\ref{sec:DCS}.
The fact that this decay channel is found here suggests that even higher-order structures might possess even more complicated and interesting stability properties.

\section{Conclusion}
\label{sec:Conclusion}
We have presented an investigation into a new family of two-dimensional solitons consisting of overlapping dark-soliton lines. In the regime where the snake instability is absent a new instability stemming from the singular point where the solitons cross can be identified, and we have discussed its behaviour for three fundamental structures: the cross soliton, the double-cross soliton, and the star soliton. For the cross soliton this instability (combined with the small system size) led to a collapse and revival of the initial state where distinct areas of identical phase connect and disconnect. The associated Bogoliubov analysis showed that this mode was well described by linear perturbation theory and that the time of its onset depends on the strength of the system's nonlinearity. The higher-order double-cross soliton and the star soliton were shown to also decay in a structured manner.

Our work shows that even though these two-dimensional soliton structures are inherently unstable, their decay process is highly structured and interesting. An obvious extension of the presented work is the generalisation to three-dimensional systems, where instabilities of a different nature could appear.

\section{Acknowledgments} 
The authors thank Jim McCann for informative discussions on the above work.


\begin{thebibliography}{0}
\bibitem{Stringari:03} L.~Pitaevskii and S.~Stringari, {\it Bose-Einstein Condensation}, Oxford University Press (2003).
\bibitem{Reinhardt:97} W.~Reinhardt and C.~Clark  J.~Phys.~B {\bf 30}, L785 (1997).
\bibitem{Busch:00} Th.~Busch and J.~R.~Anglin, Phys.~Rev.~Lett. {\bf 84}, 2298 (2000).
\bibitem{Parker:09} S.~L.~Cornish, N.~G.~Parker, A.~M.~Martin, T.~E.~Judd, R.~G.~Scott, T.~M.~Fromhold and C.~S.~Adams, Phys. D {\bf 238}, 1299 (2009).
\bibitem{Gardiner:07} A.~D.~Martin, C.~S.~Adams, and S.~A.~Gardiner, Phys.~Rev.~Lett. {\bf 98}, 020402 (2007).
\bibitem{Strecker:02} K.~E.~Strecker, G.~B.~Partridge, A.~G.~Truscott and R.~G.~Hulet, Nature (London) {\bf 417}, 150 (2002).
\bibitem{Becker:08} C.~Becker, S.~Stellmer, P.~Soltan-Panahi, S.~D\"orscher, M.~Baumert, E.~Richter, J.~Kronj\"ager, K.~Bongs and K.~Sengstock, Nat. Phys.  {\bf 4}, 496 (2008).
\bibitem{Zwierlein:13} T.~Yefsah, A.~T.~Sommer, M.~J.~H.~Ku, L.~W.~Cheuk, W.~Ji, W.~S.~Bakr \& M.~W.~Zwierlein, Nature (London) {\bf 499}, 426 (2013).
\bibitem{Anderson:01} B.~P.~Anderson, P.~C.~Haljan, C.~A.~Regal, D.~L.~Feder, L.~A.~Collins, C.~W.~Clark, and E.~A.~Cornell, Phys.~Rev.~Lett. {\bf 86}, 2926 (2001).
\bibitem{Khaykovich:02} L.~Khaykovich, F.~Schreck, G.~Ferrari, T.~Bourdel, J.~Cubizolles, L.~D.~Carr, Y.~Castin, C.~Salomon, Science {\bf 296} 1290 (2002) .
\bibitem{Burger:99} S.~Burger, K.~Bongs, S.~Dettmer, W.~Ertmer, K.~Sengstock, A.~Sanpera, G.~V.~Shlyapnikov, and M.~Lewenstein, Phys.~Rev.~Lett. {\bf 83}, 5198 (1999).
\bibitem{Denschlag:00} J.~Denschlag, J.~E.~Simsarian, D.~L.~Feder, C.~W.~Clark, L.~A.~Collins, J.~Cubizolles, L.~Deng, E.~W.~Hagley,
K.~Helmerson, W.~P.~Reinhardt, S.~L.~Rolston, B.~I.~Schneider, W.~D.~Phillips, Science {\bf 287}, 97 (2000).
\bibitem{Dutton:01} Z.~Dutton, M.~Budde, C.~Slowe and L.~V.~Hau, Science {\bf 293}, 663 (2001).
\bibitem{Stellmer:08} S.~Stellmer, C.~Becker, P.~Soltan-Panahi, E.~M.~Richter, S.~D\"{o}rscher, M.~Baumert, J.~Kronj\"{a}ger, K.~Bongs, and K.~Sengstock, Phys.~Rev.~Lett. {\bf 101}, 120406 (2008).
\bibitem{Brand:02} J.~Brand and W.~P.~Reinhardt, Phys.~Rev.~A {\bf 65}, 043612 (2002).
\bibitem{Toikka:13} L.~A.~Toikka and K.~A.~Suominen, Phys.~Rev.~A {\bf 87}, 043601 (2013).
\bibitem{Liu:05}Z.~X.~Liang, Z.~D.~Zhang, and W.~M.~Liu, Phys.~Rev.~Lett. {\bf 94}, 050402 (2005).
\bibitem{Walczak:11} P.~B.~Walczak and J.~R.~Anglin, Phys.~Rev.~A {\bf 84}, 013611 (2011).
\bibitem{Pedri:05} P.~Pedri and L.~Santos, Phys.~Rev.~Lett. {\bf 95}, 200404 (2005).
\bibitem{Baizakov:03} B.~B.~Baizakov, B.~A.~Malomed and M.~Salerno, Europhys.~Lett. {\bf 63} 642 (2003).
\bibitem{Baizakov:04} B.~B. Baizakov, B.~A.~Malomed, and M.~Salerno, Phys. Rev. A 70, 053613 (2004).
\bibitem{Xu:09} X.~Zhao, L.~Li, and Z.~Xu, Phys.~Rev.~A {\bf 79}, 043827 (2009).
\bibitem{Feder:00} D.~L.~Feder, M.~S.~Pindzola, L.~A.~Collins, B.~I.~Schneider, and C.~W.~Clark, Phys.~Rev.~A {\bf 62}, 053606 (2000).
\bibitem{Bao:03} W.~Bao, D.~Jaksch and P.~Markowich, J.~Comput.~Phys {\bf 187}, 318 (2003).
\bibitem{Fleck:76} J.~A.~Fleck, J.~R.~Morris and M.~D.~Feit, Appl.~Phys. {\bf 10} 129 (1976).
\bibitem{Bogoliubov:47} N.~N.~Bogoliubov, J.~Phys.~(USSR) {\bf 11}, 23 (1947)
\bibitem{remark1}{The Bogoliubov-de Gennes equations are numerically solved by transforming them to an eigenvalue problem using a finite-difference method. The eigenvalues and eigenfunctions are then found through standard Lanczos diagonalisation.}
\bibitem{remark2}{Even though all potential unstable modes are initially unpopulated, the numerical noise introduced during time evolution of the initial state is sufficient to populate them.}
\bibitem{Parker:03} N.~G.~Parker, N.~P.~Proukakis, M.~Leadbeater, and C.~S.~Adams, Phys.~Rev.~Lett. {\bf 90}, 220401 (2003).
\end{thebibliography}
\end{document}